\documentclass[apjl,iop]{emulateapj}
\usepackage{times,mathptmx}
\usepackage{amsmath,amssymb}
\def\substitute@command#1#2{%
 \ClassWarning{aastex}{%
  Command \protect#1\space is deprecated in aastex. 
  Using \protect#2\space instead (please fix your document). 
 }%
 #2%
}%

\slugcomment{published in \apjl, 733, L24 (2011)}

\begin{document}

\shorttitle{The end of \ion{He}{2} reionization at $z\simeq2.7$ from
\textit{HST}/COS \ion{He}{2} Ly$\alpha$ absorption spectra}
\shortauthors{Worseck et al.}

\title{The End of Helium Reionization at \lowercase{$z\simeq2.7$} Inferred from Cosmic
Variance in \textit{HST}/COS H\lowercase{e}\,{\small II} L\lowercase{y$\alpha$}
Absorption Spectra\footnotemark[$\ast$]}

\footnotetext[$\ast$]{
Based on observations made with the NASA/ESA \textit{Hubble Space Telescope}, obtained at
the Space Telescope Science Institute, which is operated by the Association of Universities
for Research in Astronomy, Inc., under NASA contract NAS5-26555
(programs 7575, 9350, 11528, 11742). Some of the data presented herein were obtained
at the W.M. Keck Observatory, which is operated as a scientific partnership among the
California Institute of Technology, the University of California and NASA; it was made
possible by the generous financial support of the W.M. Keck Foundation.
Based on observations collected at the European Organization for Astronomical Research
in the Southern Hemisphere, Chile (programs 166.A.-0106, 071.A-0066, 083.A-0421).}

\author{
G\'abor Worseck\altaffilmark{1}, J.~Xavier Prochaska\altaffilmark{1},
Matthew McQuinn\altaffilmark{2}, Aldo Dall'Aglio\altaffilmark{3},
Cora Fechner\altaffilmark{4}, Joseph~F. Hennawi\altaffilmark{5},
Dieter Reimers\altaffilmark{6}, Philipp Richter\altaffilmark{4}, Lutz Wisotzki\altaffilmark{3}
}

\altaffiltext{1}{Department of Astronomy and Astrophysics, UCO/Lick Observatory,
University of California, 1156 High Street, Santa Cruz, CA 95064, USA}
\altaffiltext{2}{Department of Astronomy, University of California, 601 Campbell Hall, Berkeley, CA 94720, USA}
\altaffiltext{3}{Astrophysikalisches Institut Potsdam, An der Sternwarte 16, 14482 Potsdam, Germany}
\altaffiltext{4}{Institut f\"{u}r Physik und Astronomie, Universit\"{a}t Potsdam, Karl-Liebknecht-Str. 24/25, 14476 Potsdam, Germany}
\altaffiltext{5}{Max-Planck-Institut f\"{u}r Astronomie, K\"{o}nigstuhl 17, 69117 Heidelberg, Germany}
\altaffiltext{6}{Hamburger Sternwarte, Universit\"{a}t Hamburg, Gojenbergsweg 112, 21029 Hamburg, Germany}
\email{gworseck@ucolick.org}

\begin{abstract}
We report on the detection of strongly varying intergalactic \ion{He}{2}
absorption in \textit{HST}/COS spectra of two $z_\mathrm{em}\simeq3$ quasars.
From our homogeneous analysis of the \ion{He}{2} absorption in these and three
archival sightlines, we find a marked increase in the mean \ion{He}{2} effective
optical depth from $\left<\tau_\mathrm{eff,HeII}\right>\simeq1$ at $z\simeq2.3$ to
$\left<\tau_\mathrm{eff,HeII}\right>\ga5$ at $z\simeq3.2$, but with a large scatter of
$2\la\tau_\mathrm{eff,HeII}\la5$ at $2.7<z<3$ on scales of $\sim10$ proper Mpc.
This scatter is primarily due to fluctuations in the \ion{He}{2} fraction and
the \ion{He}{2}-ionizing background, rather than density variations that are
probed by the co-eval \ion{H}{1} forest.
Semianalytic models of \ion{He}{2} absorption require a strong decrease in the
\ion{He}{2}-ionizing background to explain the strong increase of the absorption
at $z\ga2.7$, probably indicating \ion{He}{2} reionization was incomplete at
$z_\mathrm{reion}\ga2.7$. Likewise, recent three-dimensional numerical simulations of \ion{He}{2}
reionization qualitatively agree with the observed trend only if \ion{He}{2}
reionization completes at $z_\mathrm{reion}\simeq2.7$ or even below, as suggested
by a large $\tau_\mathrm{eff,HeII}\ga3$ in two of our five sightlines at $z<2.8$.
By doubling the sample size at $2.7\la z\la3$, our newly discovered \ion{He}{2}
sightlines for the first time probe the diversity of the second epoch of
reionization when helium became fully ionized.
\end{abstract}


\section{Introduction}

At redshifts $z\la6$, hydrogen in the intergalactic medium (IGM) is kept highly
ionized by the UV background \citep[e.g.,][]{haardt96,faucher09}, as evidenced
by the absence of strong \ion{H}{1} Ly$\alpha$ absorption in quasar sightlines
\citep{gunn65,fan06}. In contrast, the full reionization of helium
(\ion{He}{2}$\rightarrow$\ion{He}{3}) was likely delayed to
$z_\mathrm{reion}\sim3$ when quasars were sufficiently abundant to supply the
required hard $E>54.4$\,eV photons \citep{madau94,miralda-escude00}.

The $z\sim3$ \ion{H}{1} Ly$\alpha$ forest provides at best indirect evidence of
this last baryonic phase transition. The high IGM temperature measured at $z\sim3$
likely requires photoheating due to \ion{He}{2} reionization. While there is
strong evidence for a gradual reheating of the IGM over $3\la z\la4$ \citep{becker11},
evidence for a temperature peak signaling rapid \ion{He}{2} reionization is tenuous
\citep[e.g.,][]{schaye00,lidz10}. The $z\simeq3.2$ feature in the mean
\ion{H}{1} absorption \citep{theuns02b,faucher08} is unlikely due to rapid
reheating of the IGM during \ion{He}{2} reionization
\citep[][hereafter M09]{bolton09b,mcquinn09a}. Metal line systems might probe
the spectral shape of the UV background during \ion{He}{2} reionization
\citep{agafonova07,madau09}, but easily accessible metal line ratios
\citep{songaila98} are affected by metallicity variations \citep{bolton11}.

The most direct evidence for \ion{He}{2} reionization completing at
$z_\mathrm{reion}\sim3$ comes from observations of intergalactic \ion{He}{2}
Ly$\alpha$ absorption ($\lambda_\mathrm{rest}=303.78$\,\AA) toward the few
$z_\mathrm{em}\sim3$ quasars whose far-UV emission is not extinguished by
intervening \ion{H}{1} Lyman limit systems \citep{picard93,worseck11}.
\ion{He}{2} Gunn--Peterson absorption has been detected at $z>3$
\citep[e.g.,][]{jakobsen94,heap00,zheng08}, whereas the absorption becomes
patchy at $z\la3$ \citep{reimers97,reimers05}, and evolves into a \ion{He}{2}
Ly$\alpha$ forest at $z<2.7$ \citep[e.g.,][]{zheng04,fechner06}.

\begin{figure*}\centering
\includegraphics[width=0.95\textwidth]{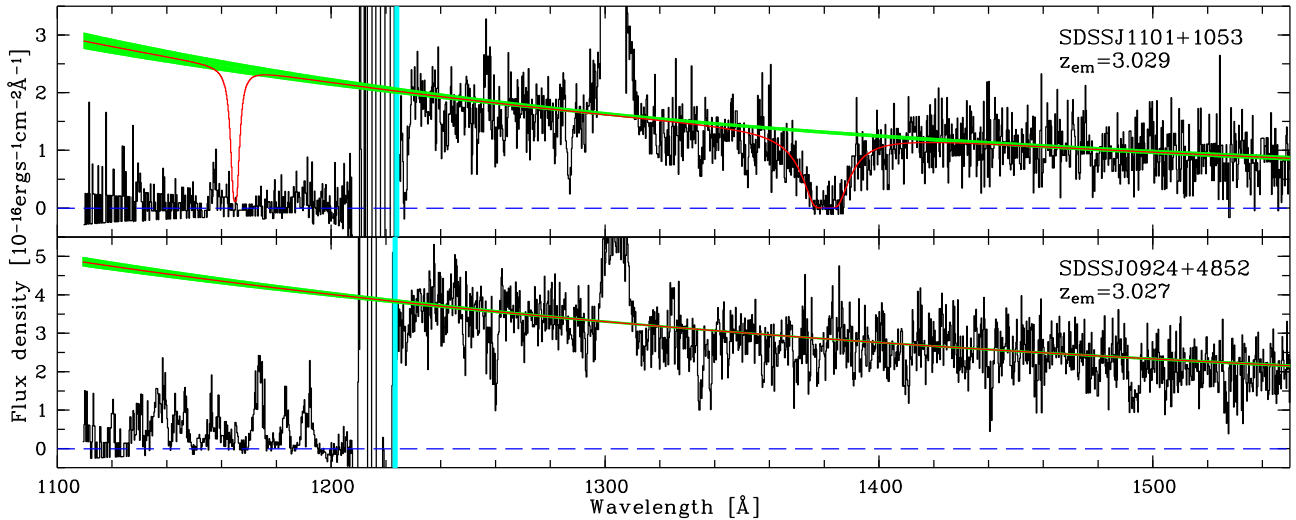}
\caption{\label{fig:he2lspc}Extinction-corrected \textit{HST}/COS far-ultraviolet
spectra of SDSS\:J1101+1053 and SDSS\:J0924+4852 displaying intergalactic
\ion{He}{2} absorption. The blue vertical bar marks \ion{He}{2} Ly$\alpha$
in the quasar rest frame, while the dashed line denotes the zero level.
The quasar proximity zones are contaminated by residuals of geocoronal Ly$\alpha$
emission. Geocoronal \ion{O}{1} emission ($\lambda\approx1300$\,\AA) has not been
subtracted. We also show the power-law fits to the quasar continua (red) and
their $1\sigma$ uncertainty estimated from Monte Carlo simulations (green).
For SDSS\:J1101+1053, we incorporate Ly$\alpha$ and Ly$\beta$ absorption of a
$\log(N_\mathrm{HI})=21.09$ damped \ion{H}{1} absorber at $z=0.1358$.}
\end{figure*}

Semianalytic \citep{gleser05,furlanetto08,furlanetto10} and numerical radiative
transfer simulations (\citealt{sokasian02}; M09) predict that \ion{He}{2}
reionization is inhomogeneous and extended. Akin to \ion{H}{1} at $z>6$
\citep{gnedin00}, \ion{He}{2} reionization is characterized by three phases:
(1) \ion{He}{3} \textquotedblleft bubble\textquotedblright\ growth around
$z_\mathrm{em}\ga4$ quasars, (2) overlap of \ion{He}{3} zones around the more
abundant quasars at $z_\mathrm{reion}\sim3$, and (3) gradual reionization of remaining
dense \ion{He}{2} regions. The large fluctuations in the \ion{He}{2} absorption
suggest that the overlap phase occurs at $z\sim3$
\citep{reimers97,smette02,jakobsen03}, but current constraints on the physics
and morphology of \ion{He}{2} reionization are limited by cosmic variance among
the handful of sightlines studied in detail.

The Cosmic Origins Spectrograph \citep[COS;][]{osterman10} on the
\textit{Hubble Space Telescope} (\textit{HST}) has the sensitivity to obtain
\ion{He}{2} absorption spectra of unprecedented quality \citep{shull10}.
Although efficient pre-selection of likely transparent sightlines with UV
photometry from the \textit{Galaxy Evolution Explorer} (\textit{GALEX}) resulted
in the discovery of 22 \ion{He}{2}-transparent quasars
\citep{syphers09b,syphers09a}, most of them are too faint for detailed studies
with COS. In this Letter, we report the discovery of two UV-bright quasars
with detected \ion{He}{2} absorption selected from our survey of the
\textit{GALEX} data set \citep{worseck11}:
\object[SDSSJ092447.36+485242.8]{SDSS\:J092447.36+485242.8} ($z_\mathrm{em}=3.027$)
and \object[SDSSJ110155.74+105302.3]{SDSS\:J110155.74+105302.3}
($z_\mathrm{em}=3.029$), hereafter SDSS\:J0924+4852 and SDSS\:J1101+1053,
respectively. Together with archival data, their diverse COS spectra constrain
the completion of \ion{He}{2} reionization to $z_\mathrm{reion}\la2.7$.
We adopt a flat cosmology with $H_0=70\,\mathrm{km}\,\mathrm{s}^{-1}\,\mathrm{Mpc}^{-1}$
and $\left(\Omega_\mathrm{m},\Omega_\Lambda\right)=\left(0.27,0.73\right)$
\citep{komatsu11}.

\section{Observations and data reduction}
\subsection{\textit{HST} Far-ultraviolet Spectra}

We obtained \textit{HST}/COS FUV spectra of SDSS\:J0924+4852 and SDSS\:J1101+1053
in course of a survey for \ion{He}{2}-transparent sightlines among the few
UV-bright $z_\mathrm{em}\sim3$ quasars detected by \textit{GALEX}
\citep[][G.~Worseck et al.\ 2011, in preparation]{worseck11}. We employed the
grating G140L in the 1105\,\AA\ setup ($\lambda\lambda$1110--2150\,\AA,
$\lambda/\Delta\lambda\sim2000$ at 1150\,\AA) at two detector settings to reduce
fixed pattern noise.

\begin{figure*}
\includegraphics[width=\textwidth]{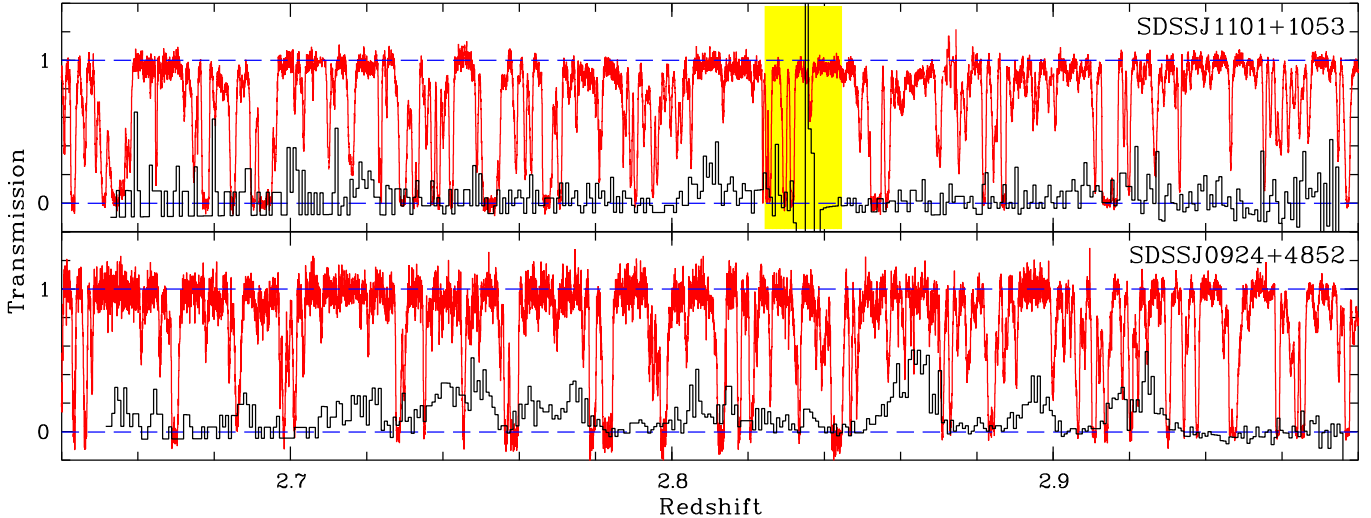}
\caption{\label{fig:h1he2trans}Optical \ion{H}{1} (red) and UV \ion{He}{2} (black)
Ly$\alpha$ absorption spectra of SDSS\:J1101+1053 ($z_\mathrm{em}=3.029$) and
SDSS\:J0924+4852 ($z_\mathrm{em}=3.027$) as a function of redshift.
The dashed lines mark the zero and the continuum level, respectively.
The shaded region in the spectrum of SDSS\:J1101+1053 corresponds to the
wavelength range impacted by Ly$\beta$ absorption of the damped \ion{H}{1}
absorber at $z=0.1358$.}
\end{figure*}

The spectra were extracted using CALCOS v2.12 and recalibrated with custom software.
Because the diffuse background is dominated by COS detector dark current,
it is well approximated by a constant over the cosmetically better detector
segment A on which the G140L spectra were recorded. A wavelength-dependent
background estimate gives very similar results, so that any
systematic background subtraction error is likely small ($\la10$\%).
Geocoronal \ion{H}{1} Ly$\alpha$ emission was subtracted. The raw counts of
individual exposures were co-added before flux calibration, thereby preserving
integer counts of our faint targets obtained in the Poisson regime. The spectra
were rebinned by a factor of three to yield approximate Nyquist
sampling ($\simeq0.24$\,\AA\,pixel$^{-1}$) at a signal-to-noise ratio (S/N) of
$\simeq2$ ($\simeq3$) per pixel for SDSS\:J1101+1053 (SDSS\:J0924+4852)
in the quasar continuum (Figure\:\ref{fig:he2lspc}).

The \textit{HST}/COS G130M and G140L data of HE\:2347$-$4342 \citep{shull10} were
retrieved from the \textit{HST} archive and reduced accordingly. For the G140L
spectrum on detector segment B, we adopted the wavelength solution by
\citet{shull10} and aligned it with the archival FUSE spectrum and the G130M
spectrum on their overlapping wavelength range. The spectra both
have an S/N$\simeq10$ per $0.24$\,\AA\ (G140L) and $0.03$\,\AA\ (G130M) pixel.

We also reanalyzed the archival \textit{HST}/STIS G140L
$\lambda/\Delta\lambda\sim1000$ spectra of Q\:0302$-$003 \citep{heap00} and
HS\:1157+3143 \citep{reimers05}. The individual exposures were extracted using
CALSTIS v2.30 and co-added, yielding a continuum S/N$\simeq5$ per
$0.6$\,\AA\ pixel.

Each extinction-corrected spectrum was normalized by fitting a power law
$f_\lambda\propto\lambda^\alpha$ on selected regions redward of
\ion{He}{2} Ly$\alpha$ in the quasar rest frame, and free of obvious emission
and absorption lines. We accounted for low-redshift Lyman limit breaks by
adopting the power law only blueward of the lowest-redshift break redward of
\ion{He}{2} Ly$\alpha$. The continuum fit was performed by maximizing the
likelihood function for the Poisson gross counts, modeling the signal part
as a power law in flux. Continuum errors were estimated by a
Monte Carlo routine, drawing Poisson deviates of the inferred continuum counts.
The $1\sigma$ statistical continuum error in the extrapolation region blueward
of \ion{He}{2} Ly$\alpha$ is $<2$\% for the high-quality spectrum of
HE\:2347$-$4342 and 5\%--10\% for the remaining quasars. Weak Lyman limit breaks
of low-redshift \ion{H}{1} absorbers present a systematic uncertainty in the
continuum in the \ion{He}{2} absorption region, although no such absorber
could be identified by its Lyman series in the low S/N data.

\subsection{Optical High-resolution \ion{H}{1} Forest Spectra}

We observed SDSS\:J0924+4852 with the Keck\:I High-Resolution Echelle
Spectrometer \citep[HIRES;][]{vogt94} for 3\,hr with the $0\farcs86$ slit
($R\sim45000$) covering the range $3300$\,\AA$\la\lambda<5930$\,\AA. The
\addtocounter{footnote}{-1}
HIRedux\footnote{\anchor{http://www.ucolick.org/~xavier/HIRedux/}{http://www.ucolick.org/\textasciitilde xavier/HIRedux/}}-reduced
spectrum has an S/N$\simeq20$ per 2.6\,$\mathrm{km}\,\mathrm{s}^{-1}$ pixel in the
\ion{H}{1} Ly$\alpha$ forest. The echelle orders were normalized by low-order
polynomials and weighted by inverse variance in their overlapping regions.

SDSS\:J1101+1053 was observed with the Very Large Telescope
(VLT) UV-Visual Echelle Spectrograph \citep[UVES;][]{dekker00} for
12.1\,hr with the 1\arcsec\ slit ($R\sim45000$) covering the range
$\lambda\lambda$3750--4980\,\AA. The spectra were reduced and normalized using
the ESO Common Pipeline
Library\footnote{\anchor{http://www.eso.org/sci/software/cpl/}{http://www.eso.org/sci/software/cpl/}},
yielding an S/N$\simeq30$ per 1.85$\,\mathrm{km}\,\mathrm{s}^{-1}$ pixel
in the Ly$\alpha$ forest.

We complemented this data set with an archival S/N$\sim100$ VLT/UVES
spectrum of HE\:2347$-$4342 \citep[e.g.,][]{worseck07}.

\section{Results}
\subsection{Detection of Intergalactic \ion{He}{2} Ly$\alpha$ Absorption}

Figure\:\ref{fig:h1he2trans} shows the cospatial \ion{H}{1} and \ion{He}{2}
Ly$\alpha$ absorption spectra of SDSS\:J1101+1053 and SDSS\:J0924+4852 as a
function of redshift. The \ion{H}{1} spectra have been corrected for
metal line absorption. At $z>2.98$ the \ion{He}{2} spectra are contaminated
by residuals of geocoronal Ly$\alpha$ emission, so that the
proximity zones of the quasars cannot be covered. Both sightlines present strong
unresolved \ion{He}{2} absorption compared to the optically thin \ion{H}{1}
forest, yet at different levels. While we see almost completely saturated
Gunn--Peterson-like absorption toward SDSS\:J1101+1053, the absorption toward
SDSS\:J0924+4852 is patchy over the whole covered redshift range. Comparing
\ion{H}{1} and \ion{He}{2} absorption, we find that \ion{H}{1} and \ion{He}{2}
do not track each other very well. Toward SDSS\:J0924+4852, a small \ion{H}{1}
void at $z\simeq2.71$ shows corresponding \ion{He}{2} transmission, but there
is stronger \ion{He}{2} transmission at $z\simeq2.86$ and $z\simeq2.92$ where
\ion{H}{1} absorption is stronger and should have caused almost complete
saturation in \ion{He}{2}. The small \ion{H}{1} void at $z\simeq2.94$ toward
SDSS\:J1101+1053 does not show obvious \ion{He}{2} transmission.
Near $z\simeq2.83$, the \ion{He}{2} absorption spectrum of SDSS\:J1101+1053 is
contaminated by \ion{H}{1} Ly$\beta$ absorption of a serendipitously discovered
damped \ion{H}{1} absorber ($z=0.1358$, $\log(N_\mathrm{HI})=21.09$).
Strong residuals after profile division render the spectral region unusable
for assessing \ion{He}{2} absorption. Neither the damped system nor the
interstellar medium causes significant metal line absorption at our low
spectral resolution.

\subsection{The Redshift Evolution of the \ion{He}{2} Effective Optical Depth}

To quantify the \ion{He}{2} absorption we computed the \ion{He}{2}
effective optical depth, $\tau_\mathrm{eff,HeII}$, in our newly discovered
sightlines and the three archival ones. We employed a novel maximum-likelihood
technique to measure $\tau_\mathrm{eff,HeII}$ in the Poisson limit in the
low-count regime of our data. The Poisson nature of the COS detector counts
becomes prominent in Figure\:\ref{fig:h1he2trans} at $z<2.7$, where single counts
are registered at a decreasing instrument response. Naive subtraction of a mean
background from these integer counts results in unphysical negative fluxes.

For a spectral segment with $n$ pixels we maximized the Poisson likelihood function
\begin{equation}
L=\prod_{j=1}^{n}\frac{\left(S_j+B_j\right)^{N_j}e^{-\left(S_j+B_j\right)}}{N_j!},
\end{equation}
with the integer number of counts per pixel $N_j$, the average background
$B_j=\mathrm{constant}$, and the unknown signal $S_j$. The signal was modeled as a
constant in transmission, $S_j=t_jC_jP_je^{-\tau_\mathrm{eff,HeII}}$ converted to
non-integer counts via the exposure time $t_j$, the flux calibration curve
$C_j$, and the power-law continuum $P_j$. Error bars (68.26\% confidence) were
calculated via integrating $L$. In case of no maximum in $L$
($\tau_\mathrm{eff,HeII}\rightarrow\infty$), we obtained the $1\sigma$ lower
limits on $\tau_\mathrm{eff,HeII}$ by refitting $\tau_\mathrm{eff,HeII}$ on mock
data generated from Poisson fluctuations of the background.

The distribution of $\tau_\mathrm{eff,HeII}$ depends on the averaging scale in
the forest. We chose a constant bin size of $\Delta z=0.04$
($\approx10$ proper Mpc at $z\sim3$) as a compromise to capture small-scale
variations in the absorption, while retaining enough sensitivity to measure high
effective optical depths against the Poisson detector background. Small bin size
variations do not change our results.
For all five sightlines, we adopted identical redshift bins without focusing on
particular features. Proximity zones, geocoronal emission, and the
damped \ion{H}{1} absorber toward SDSS\:J1101+1053 were omitted from our analysis.
For HE\:2347$-$4342, we adopted $\tau_\mathrm{eff,HeII}$ from the G130M spectrum
for completely covered redshift bins.

\begin{figure}
\includegraphics[width=3.39in]{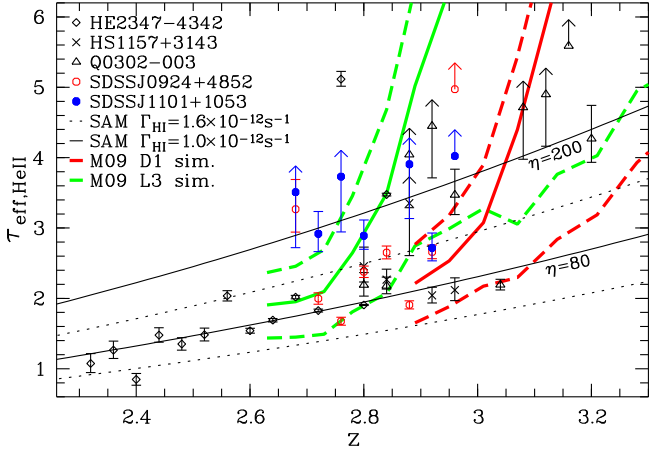}
\caption{\label{fig:taueffz}\ion{He}{2} effective optical depth
$\tau_\mathrm{eff,HeII}$ as a function of redshift $z$ for five \ion{He}{2}
sightlines in identical redshift bins of $\Delta z=0.04$
($\approx10$ proper Mpc at $z\sim3$). Double-sided error bars
are 68.26\% confidence, whereas single-sided error bars and lower limits are
$84.13$\% confidence ($1\sigma$). Overplotted are predictions of
$\tau_\mathrm{eff,HeII}(z)$ from our optically thin semianalytic IGM models
and results from radiative transfer simulations of \ion{He}{2} reionization
by \citealt{mcquinn09a} (solid: mean $\tau_\mathrm{eff,HeII}$, dashed: $1\sigma$
deviation).}
\end{figure}

Figure\:\ref{fig:taueffz} and Table\:\ref{tab:taueff} present
$\tau_\mathrm{eff,HeII}(z)$ for the five sightlines in the redshift range
$2.32\le z\le3.20$. The \ion{He}{2} effective optical depth evolves strongly from
$\tau_\mathrm{eff,HeII}\simeq1$ at $z\simeq 2.3$ to $\tau_\mathrm{eff,HeII}\ga5$
at $z\simeq3.2$, although cosmic variance might play a role at the lowest and highest
redshifts (probed by one sightline each).
The effective optical depth in the HE\:2347$-$4342 sightline rather smoothly increases
with redshift until $z\simeq2.7$. Between $z\simeq2.7$ and $z\simeq3$, we observe
a large scatter in $\tau_\mathrm{eff,HeII}$. In four of our five sightlines,
approximately half of the data points continue the smooth trend from lower
redshifts, whereas the remaining ones are significantly higher. For example,
the high-quality spectrum of HE\:2347$-$4342 shows $\tau_\mathrm{eff,HeII}\simeq5.1$
at $z=2.76$. In the sightline toward SDSS\:J1101+1053, strong absorption with
$\tau_\mathrm{eff,HeII}\ga3$ occurs almost everywhere, whereas
$\tau_\mathrm{eff,HeII}$ mostly fluctuates around $\tau_\mathrm{eff,HeII}\sim2$
toward SDSS\:J0924+4852. Apart from the known \ion{He}{2} void at $z=3.05$
\citep[e.g.,][]{jakobsen03}, the Q\:0302$-$003 sightline shows
$\tau_\mathrm{eff,HeII}\ga5$ at $z>3$.

\begin{figure}
\includegraphics[width=3.39in]{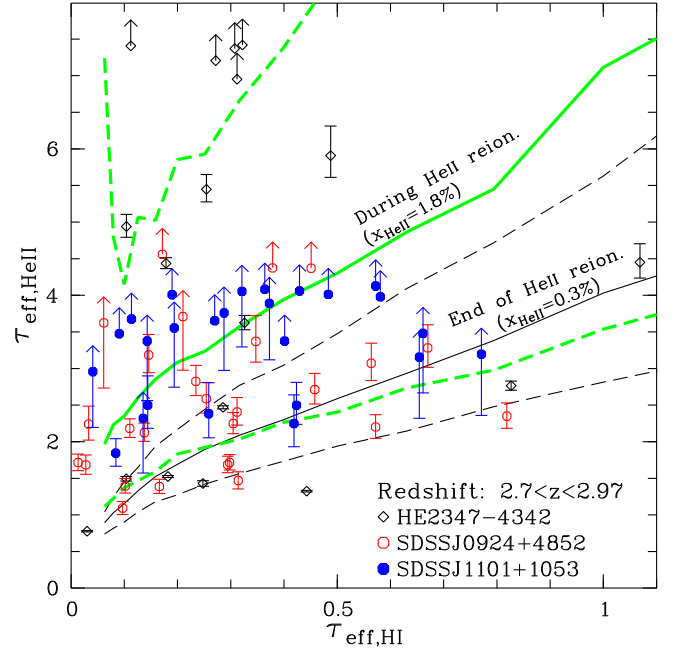}
\caption{\label{fig:tauh1he2}\ion{He}{2} effective optical depth
$\tau_\mathrm{eff,HeII}$ as a function of the corresponding \ion{H}{1} effective
optical depth $\tau_\mathrm{eff,HI}$ in three \ion{He}{2} sightlines over $2.7<z<2.97$
on $\Delta z=0.01$ bins ($\simeq2.8$ proper Mpc at $z=2.8$).
Error bars are $1\sigma$. Overplotted are the median (solid) and the 15.87 and 84.13
percentiles (dashed) of the $\tau_\mathrm{eff,HeII}$
distribution at a given $\tau_\mathrm{eff,HI}$ from the M09 simulations for
ongoing (volume-averaged \ion{He}{2} fraction $x_\mathrm{HeII}=1.8$\%, green)
and almost complete ($x_\mathrm{HeII}=0.3$\%, black) \ion{He}{2} reionization.}
\end{figure}

Small-scale density variations might contribute to the effective optical depth
variations. We investigated this by comparing the \ion{He}{2} absorption with the
cospatial \ion{H}{1} Ly$\alpha$ absorption on a small scale $\Delta z=0.01$
($\simeq2.8$ proper Mpc at $z=2.8$). We focused on the redshift range $2.7<z<2.97$
where the scatter and the data coverage is largest. Figure\:\ref{fig:tauh1he2}
presents our measurements toward the three quasars with \ion{H}{1} and
\ion{He}{2} data. From the absence of a clear correlation between the measured
\ion{H}{1} and \ion{He}{2} absorption we conclude that the $\tau_\mathrm{eff,HeII}$
fluctuations cannot be due to IGM density variations alone.

\begin{deluxetable}{lll}
\tabletypesize{\scriptsize}
\tablewidth{2.0in}
\tablecaption{\label{tab:taueff}\ion{He}{2} Effective Optical Depths ($\Delta z=0.04$)}
\tablehead{
\colhead{Quasar}&\colhead{$z$}&\colhead{$\tau_\mathrm{eff,HeII}$}}
\startdata
HE\:2347$-$4342		&$2.32$ &$1.08^{+0.14}_{-0.13}$\\
			&$2.36$ &$1.26^{+0.13}_{-0.12}$\\
			&$2.40$ &$0.85^{+0.09}_{-0.08}$\\
			&$2.44$ &$1.48^{+0.11}_{-0.10}$\\
			&$2.48$ &$1.35^{+0.09}_{-0.09}$\\
			&$2.52$ &$1.48^{+0.09}_{-0.09}$\\
			&$2.56$ &$2.04^{+0.07}_{-0.07}$\\
			&$2.60$ &$1.54^{+0.04}_{-0.04}$\\
			&$2.64$ &$1.69^{+0.02}_{-0.02}$\\
			&$2.68$ &$2.02^{+0.03}_{-0.03}$\\
			&$2.72$ &$1.83^{+0.02}_{-0.02}$\\
			&$2.76$ &$5.12^{+0.11}_{-0.10}$\\
			&$2.80$ &$1.91^{+0.01}_{-0.01}$\\
			&$2.84$ &$3.48^{+0.02}_{-0.02}$\\
HS\:1157+3143		&$2.80$	&$2.46^{+0.27}_{-0.22}$\\
			&$2.84$	&$2.27^{+0.15}_{-0.13}$\\
			&$2.88$	&$3.35^{+\infty}_{-0.75}$\\
			&$2.92$	&$2.04^{+0.12}_{-0.11}$\\
			&$2.96$	&$2.12^{+0.17}_{-0.15}$\\
SDSS\:J0924+4852	&$2.68$	&$3.27^{+0.42}_{-0.33}$\\
			&$2.72$	&$2.00^{+0.08}_{-0.08}$\\
			&$2.76$	&$1.68^{+0.06}_{-0.06}$\\
			&$2.80$	&$2.38^{+0.08}_{-0.08}$\\
			&$2.84$	&$2.65^{+0.09}_{-0.09}$\\
			&$2.88$	&$1.91^{+0.06}_{-0.06}$\\
			&$2.92$	&$2.66^{+0.10}_{-0.10}$\\
			&$2.96$	&$4.97^{+\infty}_{-0.00}$\\
SDSS\:J1101+1053	&$2.68$	&$3.51^{+\infty}_{-0.79}$\\
			&$2.72$	&$2.92^{+0.32}_{-0.26}$\\
			&$2.76$	&$3.73^{+\infty}_{-0.79}$\\
			&$2.80$	&$2.89^{+0.22}_{-0.19}$\\
			&$2.88$	&$3.91^{+\infty}_{-0.77}$\\
			&$2.92$	&$2.72^{+0.21}_{-0.18}$\\
			&$2.96$	&$4.02^{+\infty}_{-0.00}$\\
Q\:0302$-$003		&$2.80$	&$2.19^{+0.18}_{-0.16}$\\
			&$2.84$	&$2.18^{+0.12}_{-0.11}$\\
			&$2.88$	&$4.04^{+\infty}_{-0.76}$\\
			&$2.92$	&$4.45^{+\infty}_{-0.74}$\\
			&$2.96$	&$3.47^{+0.37}_{-0.28}$\\
			&$3.04$	&$2.19^{+0.08}_{-0.08}$\\
			&$3.08$	&$4.71^{+\infty}_{-0.74}$\\
			&$3.12$	&$4.90^{+\infty}_{-0.74}$\\
			&$3.16$	&$5.59^{+\infty}_{-0.00}$\\
			&$3.20$	&$4.27^{+0.47}_{-0.34}$
\enddata
\end{deluxetable}

\section{Discussion}

\subsection{Comparison to Models}

To further explore the results in Figure\:\ref{fig:taueffz} we constructed a
simple semianalytic model for \ion{He}{2} absorption in a reionized IGM. For an
IGM highly photoionized in H and He that follows the temperature--density
relation $T\left(\Delta=\rho/\bar{\rho}\right)=T_0\Delta^{\gamma-1}$
\citep{hui97}, the \ion{H}{1} optical depth is
\begin{eqnarray}\label{eq:fgpa}
\tau_\mathrm{HI}&\simeq&0.612\left(\frac{T_0}{20,000\mathrm{K}}\right)^{-0.724}
\left(\frac{\Gamma_\mathrm{HI}}{10^{-12}\mathrm{s}^{-1}}\right)^{-1}\\\nonumber
&\times&\Delta^{2-0.724(\gamma-1)}\left(\frac{1+z}{4}\right)^{4.5},
\end{eqnarray}
also known as the fluctuating Gunn--Peterson approximation
\citep[e.g.,][]{weinberg97b}. \ion{H}{1} and \ion{He}{2} trace the same cosmic
densities, so
\begin{equation}
\tau_\mathrm{HeII}\simeq0.112\frac{\Gamma_\mathrm{HI}}{\Gamma_\mathrm{HeII}}\tau_\mathrm{HI}\simeq\frac{\eta}{4}\tau_\mathrm{HI}
\end{equation}
in a reionized IGM with an \ion{H}{1} (\ion{He}{2}) photoionization rate
$\Gamma_\mathrm{HI}$ ($\Gamma_\mathrm{HeII}$). We have approximated number
densities as column densities with the column density ratio
$\eta=N_\mathrm{HeII}/N_\mathrm{HI}$ commonly measured in \ion{He}{2} forest
spectra. The \ion{He}{2} effective optical depth can be written as
\begin{equation}
\tau_\mathrm{eff,HeII}=-\ln{\left[\int_0^\infty e^{-\frac{\eta}{4}\tau_\mathrm{HI}}P\left(\tau_\mathrm{HI}\right)d\tau_\mathrm{HI}\right]}
\end{equation}
with the \ion{H}{1} optical depth probability distribution function
$P\left(\tau_\mathrm{HI}\right)=P\left(\Delta\right)\left| d\Delta/d\tau_\mathrm{HI}\right|$.
With Equation\:(\ref{eq:fgpa}) and the overdensity probability distribution
$P\left(\Delta\right)$ from simulations by \citet{bolton09c}, $\tau_\mathrm{eff,HeII}$ is a function
of the temperature--density relation ($T_0,\gamma$) and the ionization conditions
($\Gamma_\mathrm{HI},\eta$). We adopt $T_0=15,000$\,K and the post-reionization
asymptotic value $\gamma=1.5$ \citep{hui97},
$\Gamma_\mathrm{HI}=10^{-12}\,\mathrm{s}^{-1}$ \citep{bolton05} and $\eta=80$
\citep{fechner07}.

The lower solid curve in Figure\:\ref{fig:taueffz} shows the resulting
$\tau_\mathrm{eff,HeII}(z)$ model. It matches surprisingly well to the measured
values at $z<2.7$. Part of the data follows this relation until $z\sim3$,
whereas the rest significantly departs from it to larger $\tau_\mathrm{eff,HeII}$
values. The amplitude of $\tau_\mathrm{eff,HeII}(z)$ mostly depends on the
ionization conditions, indicated in Figure\:\ref{fig:taueffz} by a higher
$\Gamma_\mathrm{HI}$ \citep{dallaglio08} or a softer UV background (higher $\eta$).
However, the steep evolution of $\tau_\mathrm{eff,HeII}$ cannot be matched
unless there is significant redshift evolution in the model parameters.
A change in the temperature--density relation seems implausible, since a
flattening would cause $\tau_\mathrm{eff,HeII}$ to decrease, and a rise
in $\tau_\mathrm{eff,HeII}$ would require unreasonably low IGM temperatures.
Since $\Gamma_\mathrm{HI}$ and $\eta$ are tied, the steep increase of
$\tau_\mathrm{eff,HeII}$ might suggest a strong decrease of
$\Gamma_\mathrm{HeII}$ at $z>2.7$ related to a drop in the mean free path of
\ion{He}{2}-reionizing photons if the quasar emissivity is constant. While this
alone likely indicates that \ion{He}{2} reionization is occurring, the breakdown
of the high-ionization limit of helium could simply invalidate our semianalytic
approach at $z>2.7$.

To describe the redshift evolution of the effective optical depth during
reionization, we calculated $\tau_\mathrm{eff,HeII}(z)$ from numerical models of
\ion{He}{2} reionization (M09). We focused on their D1 and L3 simulations, which
have different reionization histories due to the filtering of UV radiation by
dense absorbers. Run L3 supplements the D1 simulation with absorbers that may not
be resolved and which delay completion of reionization slightly from
$z_\mathrm{reion}\simeq3$ (D1) to $z_\mathrm{reion}\simeq2.7$ (L3). From each
simulation snapshot we computed 1000 skewers across the box, each providing
$\sim5$ samples of $\tau_\mathrm{eff,HeII}$ on the same scale as our measurements
($\Delta z=0.04$).

The thick curves in Figure\:\ref{fig:taueffz} show the resulting mean
$\tau_\mathrm{eff,HeII}$ (solid) and its $1\sigma$ scatter (dashed). Both models
predict a strong increase in the \ion{He}{2} absorption as their respective
volume-averaged \ion{He}{2} fraction rises, in contrast to our optically thin
models. On our chosen scale $\Delta z=0.04$, the models exhibit large
fluctuations in $\tau_\mathrm{eff,HeII}$ due to cosmic variance in the
\ion{He}{2} reionization histories. At the end of \ion{He}{2} reionization,
both models show a characteristic turnover in $\tau_\mathrm{eff,HeII}$ to
approximately our favored optically thin model of the post-reionization IGM.
The data are largely inconsistent with the predictions from model D1 of M09 due
to the large excess toward high $\tau_\mathrm{eff,HeII}$ at $z<3$.
Model L3, meanwhile, also does not reproduce the data perfectly, since the largest
$\tau_\mathrm{eff,HeII}$ values at $z\simeq$2.7--2.8 and the lowest values at
$z\simeq$2.9--3 cannot be easily accommodated.

In Figure\:\ref{fig:tauh1he2}, we compare the measured \ion{H}{1} and \ion{He}{2}
effective optical depths to $\sim20,000$ mock samples of two snapshots of the L3
simulation by M09, obtained on the same scale $\Delta z=0.01$ and rescaled to
a common redshift $z\simeq2.8$.
With respective volume-averaged \ion{He}{2} fractions of $x_\mathrm{HeII}=1.8$\%
and $x_\mathrm{HeII}=0.3$\%, the simulation outputs trace the end stages of
\ion{He}{2} reionization. In the mock spectra, helium and hydrogen absorptions are
correlated due to the underlying density field and the emerging
\ion{He}{2}-ionizing background. The scatter in this relation is primarily due to
fluctuations in the \ion{He}{2} fraction and the UV background, which decrease
as \ion{He}{2} reionization proceeds.
\ion{H}{1} voids (low $\tau_\mathrm{eff,HI}$) without nearby UV sources contain
large amounts of \ion{He}{2} if reionization is still ongoing. The largest
inferred effective optical depths ($\tau_\mathrm{eff,HeII}\ga7$) toward
HE\:2347$-$4342 indicate that the \ion{He}{2} fraction is still $>2$\%. 
The SDSS\:J1101+1053 sightline is mostly consistent with the scenario of still
incomplete \ion{He}{2} reionization, whereas parts of the SDSS\:J0924+4842
sightline are better matched with a reionized IGM, which, however, does not contain
the strong fluctuations seen in the data.
Thus, while data and the M09 models suggest that we see the end stages of \ion{He}{2}
reionization, the coexistence of patches with both large and small \ion{He}{2}
fractions at moderate densities is difficult to explain. A more rapid increase in the
quasar emissivity, a short quasar lifetime or quasar anisotropy
\citep[e.g.,][]{hennawi07} might amplify the modeled fluctuations to the required level.

\subsection{Implications}

Our newly discovered quasar sightlines with detected \ion{He}{2} Ly$\alpha$
absorption highlight the diversity in \ion{He}{2} absorption near the epoch of
\ion{He}{2} reionization, irrespective of the gas density probed by the coeval
\ion{H}{1} Ly$\alpha$ forest. The alternating voids and troughs toward
SDSS\:J0924+4852 are expected at the very end of \ion{He}{2} reionization, in
agreement with current numerical modeling \citep{mcquinn09a}. In contrast, the
two long ($\ga10$Mpc) \ion{He}{2} absorption troughs toward HE\:2347$-$4342
indicate that \ion{He}{2} reionization is still incomplete. In agreement with
\citet{furlanetto10} and \citet{shull10} we find that the large
$\tau_\mathrm{eff,HeII}\simeq5.1$ at $z\simeq2.76$ in the HE\:2347$-$4342
sightline is inconsistent with the reionized IGM in the L3 simulation by
\citet{mcquinn09a} at $\sim98$\% confidence.
Extended trough-like absorption in the SDSS\:J1101+1053 sightline
(Figure\:\ref{fig:h1he2trans}) supports the view that \ion{He}{2} reionization
is delayed to $z_\mathrm{reion}\simeq2.7$ or even below. 

When measured on a fixed spatial scale, the redshift evolution of the \ion{He}{2}
effective optical depth is a powerful probe of \ion{He}{2} reionization.
The marked upturn in $\tau_\mathrm{eff,HeII}$ at $z\ga2.7$ indicates a strong
decrease in the \ion{He}{2}-ionizing background coupled to the mean free path of
\ion{He}{2}-ionizing photons, or a strongly increasing \ion{He}{2} abundance.
Both of these suggest that \ion{He}{2} reionization is occurring at $z\ga2.7$.
Similar conclusions were obtained by \citet{dixon09}, who inferred a dramatic
decrease in the \ion{He}{2}-ionizing background or a decrease in the mean free
path at $z>2.8$. However, the sparse data then available provided only weak
constraints on the epoch of \ion{He}{2} reionization. 

By combining our data with three-dimensional numerical radiative transfer simulations
\citep{mcquinn09a}, we constrain the end of the \ion{He}{2} reionization epoch
to $z_\mathrm{reion}\la2.7$. The steep increase in $\tau_\mathrm{eff,HeII}$
at $z>2.8$ is well reproduced by models of significantly delayed \ion{He}{2}
reionization ($z_\mathrm{reion}\simeq2.7$) driven by quasars, whereas the
strong absorption at $z<3$ rules out models with \ion{He}{2} reionization
completing at $z_\mathrm{reion}\ga3$. The coexistence of regions with strong
and weak ionization at $2.7<z<3.0$ suggests that our \textit{HST}/COS spectra
probe the end stages of \ion{He}{2} reionization. We speculate that the variance
in the data might be explained with more sophisticated modeling of quasars in
\ion{He}{2} reionization simulations (e.g., their number density, spectral
energy distribution, lifetime, anisotropy). Thus, \ion{He}{2} reionization might
yield unique insight into the quasar phenomenon.

\acknowledgments
We acknowledge support by an NSF CAREER grant (AST-0548180) and by NSF grant
AST-0908910. Support for program 11742 was provided by NASA through a grant from
the Space Telescope Science Institute, which is operated by the Association of
Universities for Research in Astronomy, Inc., under NASA contract NAS5-26555.


\bibliographystyle{apj}
\bibliography{he2cos}

\end{document}